      [1999/12/01 v1.4c Il Nuovo Cimento]
\documentclass{cimento}
\usepackage{graphicx}  
\usepackage{epsfig} 
\title{Identification of Relativistic Effects in X-Ray Astrophysical Sources}
\author{Costantino Sigismondi\from{ins:x}}
\instlist{\inst{ins:x} ICRA, International Center for Relativistic 
Astrophysics, and University of Rome La Sapienza, Piazzale Aldo Moro 5 00185 Roma, Italy.}
\PACSes{\PACSit{98.70.Q}{X-ray Astronomy}
\PACSit{95.30.S}{General Relativity}
\PACSit{97.60.Cf}{Black Holes}}
\begin{document}

\maketitle

\begin{abstract}
Precise knowledge of the features of several X-ray sources from compact binary sources to galactic nuclei is gradually unveiling the physical parameters of relativistic neutron stars and black holes therein. Recent developments in theory, astrophysical models and experiments are reviewed.
\end{abstract}

\section{Introduction}

In the last years appeared several works dealing with interpretations of X-ray sources' observations, mainly made with Rossi XTE satellite \cite{ref:sigi}, and following the main streams of the debate on the identification of Lense-Thirring frequency in QPOs \cite{ref:stella} or the classification of ``Z and atoll sources'' \cite{ref:vdk}.

\section{Observational phenomena of ``Z'' and ``atoll'' sources}

The presence of neutron stars in low-mass X-ray binaries (LMXBs) is revealed by the occurrence of X-ray bursts or pulsations in these systems. The brigthest LMXBs, as Cygnus X-2 (observed with BeppoSAX  \cite{ref:Piraino}) and Sco X-1, were called ``Z sources'' because of the Z pattern in the X-ray color-color diagram: horizontal branch; top stroke of the Z; normal branch (middle stroke) and flaring branch (end stroke). The ``Z Sources'' were thought to have a mass accretion rate near the Eddington limit. \cite{ref:Fortner}
Thanks to the microsecond accuracy of the Rossi X-ray Timing Explorer
launched in 1995 it has been possible to progress largely in the interpretation of the nature of LMXBs \cite{ref:Vaughan} and now the entire catalogue of observations is available online. \cite{ref:XTE}

In figure 1 there is the explanation about ``Z sources'' denomination.

Hasinger and Van der Klis \cite{ref:Hasinger} found in EXOSAT data (time resolution of milliseconds) the evidence of a new class of sources with a different pattern of correlated timing and spectral behaviour. Because of their fragmented color-color diagram they called them ``atoll sources''.
Z sources tend to be more luminous than atoll sources. \cite{ref:Klis90}
The patterns which atoll sources trace in X-ray color-color diagrams are similar to the shapes for which the more-luminous Z sources are named.
\begin{figure} 
\centering \includegraphics[height=1.0\columnwidth, angle=-90]{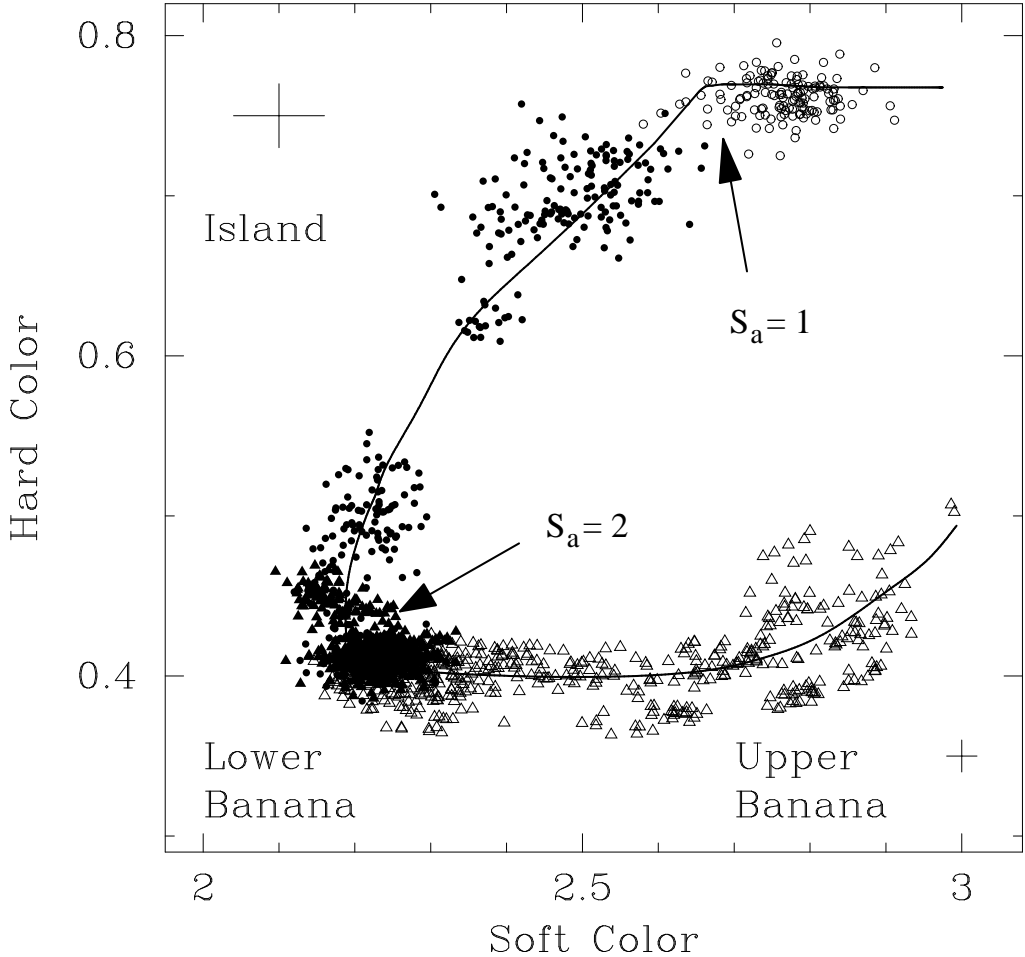} 
\caption[]{X-ray color-color diagram of the atoll source 4U\,1608$-$52. Mass accretion
rate is inferred to increase counterclockwise, approximately along the
drawn curve. kHz QPO detections are indicated with filled
symbols (credit: Van der Klis  \cite{ref:sigi}). A Z source traces a complete Z on this diagram.}
\label{fig:uno} 
\end{figure}

\subsection{Quasi-periodic oscillators: QPOs}

Quasi-periodic oscillations QPOs have been discovered also in atoll
sources with Rossi XTE. 
Analyses of those sources have been published until now, and compared with theoretical models.
Timing properties of the bursting atoll source 4U 1728-34 have been studied
as a function of its position in the X-ray color-color diagram. \cite{ref:DiSalvo} A detailed discussion of the different components of the signal (as very low frequency noise below 1 Hz, low-frequency QPO between 20-40 Hz) is also present.

\begin{figure} 
\centering \includegraphics[height=1.0\columnwidth, angle=0]{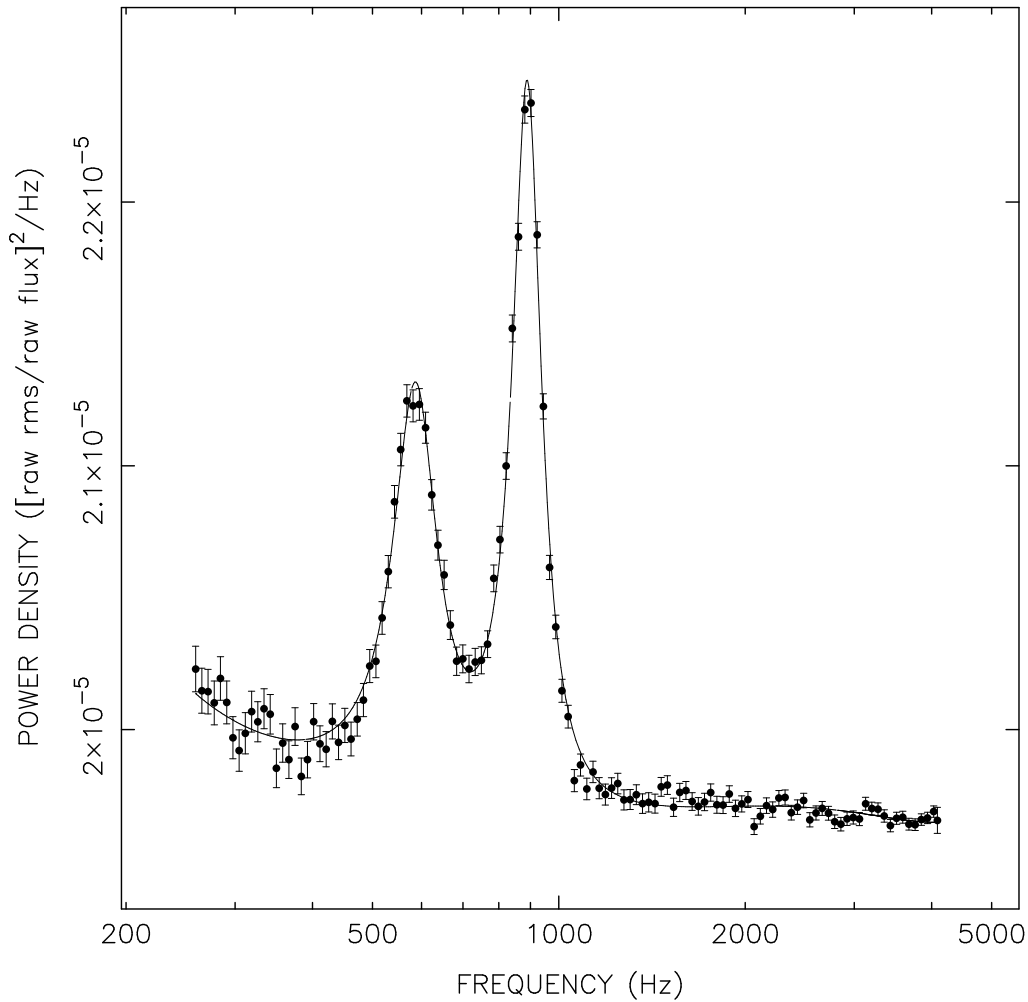} 
\caption[]{Twin kHz peaks in Sco\,X-1(credit: Van der Klis  \cite{ref:sigi}). It is a Z source showing kilohertz QPOs.}
\label{fig:due} 
\end{figure}

\section{Different models of QPO}

Several models in addition to those related to Lense-Thirring precession \cite{ref:stella,ref:Markovich}  have been used to explain such observational features: QPOs due to orbital motion of clumps \cite{ref:Karas1}; epicyclic motions in the accretion disk \cite{ref:Torkelsson}; transonic disk accretion flows \cite{ref:Lee,ref:Lai2}, stellar oblateness \cite{ref:Stella5}, photospheric radius expansion \cite{ref:Muno2}, and beaming oscillations \cite{ref:Miller2}. Homan et al. \cite{ref:Homan} compare the behavior of GX 17+2 and other Z sources with that of black hole sources and consider the possibility that the mass accretion rate might not be driving force behind all spectral and variability changes. 

\subsection{Transition layer model}

Wu \cite{ref:Wu} compared the theoretical predictions for neutron stars X-ray binaries of the transition layer model with some observational features of QPOs and suggested that similar transition layer oscillations may be responsible for the 
observed QPOs in accretion-powered millisecond X-ray pulsars and 
galactic black hole candidates. 

\subsection{The role of stellar magnetic field}
The inner regions of accretion disks of weakly magnetized neutron stars 
are affected by general relativistic gravity and stellar magnetic 
fields. Lai \cite{ref:Lai} posed an upper limit to the surface magnetic field $B_o$ can 
be obtained, i.e.,  $B_o \leq 3\cdot10^7(\dot{M} _{17}/ \beta )^{1/2} ~G$, where 
$\dot{M} _{17}=\dot{M} /(10^{17}~ g\cdot s^{-1})$, in order to produce kHz orbital 
frequency at the sonic radius. 
An accretion disk around a rotating magnetized star is subjected to 
magnetic torques that induce disk warping and precession. General studies on thin disks have been recently published \cite{ref:Kato}, and around weakly magnetized neutron stars. \cite{ref:Lee2} These torques 
arise generically from interactions between the stellar field and 
induced surface currents on the disk. Applying these new effects to 
weakly magnetized ($B\sim 10^7-10^9~ G$) neutron stars in 
low-mass X-ray binaries, Shirakawa and Lai \cite{ref:Shirakawa} studied the global hydrodynamical 
warping/precession modes of the disk under the combined influences of 
relativistic frame dragging, classical precession due to the oblateness 
of the neutron star, and the magnetic torques. 
The modes 
are confined to the inner region of the disk and have frequencies equal 
to $0.3\div0.95$ (depending on the mass accretion rate $\dot{M}$) times the Lense-Thirring frequency.
The magnetically driven precession is retrograde 
(opposite to the Lense-Thirring precession).
This may account for several observed features of 
low-frequency (10-60 Hz) quasi-periodic oscillations in 
LMXBs.
Interactions with neutron stars magnetic fields have been invoked in other papers \cite{ref:Campana, ref:Lee2}
The {\it magnetospheric beat-frequency}  \cite{ref:Psaltis} QPO is the model proposed to explain the signal of 4U 1735-44 and Z sources in Wijnands et al. \cite{ref:Wijnands}.

\section{Lense-Thirring effect on warped accretion disks}

Among the effects around a rapidly rotating compact object is  
the Bardeen-Petterson effect \cite{ref:Bardeen} which
causes a tilted accretion disk to warp into the equatorial plane of the rotating body. Viscous forces cause the accretion flow to divide into 
two distinct regions-an inner aligned accretion disk and an outer tilted 
accretion disk. The transition between these two regimes occurs at a 
characteristic radius that depends on the mass and angular momentum of 
the central object and possibly on the accretion rate through the disk. 
Fragile et al.  \cite{ref:Fragile} propose that the accreting material passing through the transition region 
may generate quasi-periodic brightness oscillations such as have been 
observed in a number of X-ray binaries. QPO frequency range predicted by this 
model is consistent with observed QPO frequencies in both black hole and 
neutron star low-mass X-ray binaries. Armitage and Natarajan \cite{ref:Armitage} studied the Lense-Thirring and Bardeen-Petterson effect acting upon warped accretion disks.
Sibgatullin gave compact approximation formulas for the nodal precession frequency of the marginally stable circular orbits around black holes  \cite{ref:Sibga1} and around rapidly rotating neutron stars.  \cite{ref:Sibga2}

\section{Gravitomagnetism around neutron stars and black holes}

The main processes that produce the variability are the same for black holes 
and neutron stars  \cite{ref:Belloni}. The distinction among them is suggested studying Fourier spectra at high frequencies  \cite{ref:Sunyaev}.
Galloway et al.  \cite{ref:Galloway} studying the discovery of a highly coherent oscillation in a type I X-ray burst observed from 4U 1916-053 by the Rossi XTE guessed the
neutron star spin period to be around 3.7 milliseconds. Begelman  \cite{ref:Begelman} also
agrees that observational techniques and theoretical ideas are enabling us to find black holes and measure their masses with increasing precision, and we may soon be able to measure black hole's spins. 

\section{Galactic black holes}

Melia et al. \cite{ref:Melia} suggested that 
submillimeter timing observations could yield a signal corresponding to 
the period $P_0$ of the marginally stable orbit and therefore 
point directly to the black hole's spin $\vec{s}$. Sgr A*'s mass is now known to be $2.6 \pm 0.2 \times 10^6~ M_{\odot}$ (an unusually 
accurate value for supermassive black hole candidates), for which 2.7 minutes $\leq P_0 \leq$ 36 minutes, depending on the value of $\vec{s}$ and whether the Keplerian flow is prograde or retrograde. A Schwarzschild 
black hole ($\vec{s}=0$) should have $P_0\sim$ 20 minutes. The 
identification of the orbital frequency with the innermost stable 
circular orbit is made feasible by the transition from optically thick 
to thin emission at submillimeter wavelengths. By 
analogy with low-mass X-ray binaries and Galactic black hole candidates, 
Sgr A* may also display quasi-periodic oscillations. Other authors studied matter flows onto galactic black holes in the same perspective of X-ray binaries or gravitomagnetism \cite{ref:Natarajan}.
 Alternative derivations of Lense-Thirring effect \cite{ref:Iorio}, as well as some derivations of the ``clock effect'' related to the non static part of the metric in General Relativity \cite{ref:Iorio1,ref:Iorio3,ref:Iorio5,ref:Bini} like time delay in gravitational lensing due to the rotation of a foreground galaxy \cite{ref:Ciufolini1,ref:Ciufolini2} have been proposed.

\section{Experiments around Earth}

Waiting for GP-B data, the satellite has been launched in april 2004\cite{ref:GPB}, and considering all space experiments already done on this topic \cite{ref:sigi}, the dramatic improvement of this field it is expected from two years after the launch (end of 2006). 
C. L\"{a}mmerzahl coedited with C.W.F. Everitt and F. W. Hehl a volume dedicated to tests of General Relativity in Space \cite{ref:Lam}.

About measurements of Lense-Thirring effect around Earth there are the studies on the influence of Earth tides \cite{ref:Iorio1,ref:Iorio3}, Earth's penumbra \cite{ref:Vespe}, non gravitational perturbations modeling \cite{ref:Lucchesi} and the use of clocks. \cite{ref:Lichtenegger}
Pascual-S{\'a}nchez \cite{ref:Pascual2} proposes a measurement of gravitomagnetism in a terrestrial laboratory.

\end{document}